\documentstyle[epsfig,]{aipproc}
\newcommand{\cd}{\cdot}

\renewcommand{\b}{\beta}
\newcommand{\de}{\delta}
\newcommand{\De}{\Delta}
\newcommand{\ep}{\epsilon}

\newcommand{\La}{\Lambda}
\newcommand{\la}{\lambda}
\newcommand{\Om}{\Omega}

\newcommand{\si}{\sigma}
\newcommand{\Si}{\Sigma}

\newcommand{\ra}{\rightarrow}
\newcommand{\bk}{{\bf k}}

\newcommand{\apj}{ Astrophys. J.}
\renewcommand{\baselinestretch}{1.01}
\newcommand{\be}{\begin{equation}}
\newcommand{\ee}{\end{equation}}
\newcommand{\bea}{\begin{eqnarray}}
\newcommand{\eea}{\end{eqnarray}}
\newcommand{\bean}{\begin{eqnarray*}}
\newcommand{\eean}{\end{eqnarray*}}
\newcommand{\dd}{\partial}

\newcommand{\gsim}{\stackrel{>}{\sim}}

\newcommand{\Mpc}{{\rm Mpc}}
\renewcommand{\r}{\right}
\renewcommand{\l}{\left}
\title{Cosmic Microwave Background Anisotropies
 from Global Texture}
\author{R. Durrer, M. Kunz and A. Melchiorri}
\address{D\'epartement de Physique Th\'eorique,
         Universit\'e de Gen\`eve,
        24 quai Ernest Ansermet,\\ CH-1211 Gen\`eve 4, Switzerland}
\begin{document}
\maketitle
\begin{abstract}
We investigate the global texture model of structure formation in
cosmogonies with non-zero cosmological constant for different
values of the Hubble parameter. We find that the absence of
significant acoustic peaks and little power on large scales
are robust predictions of these models. However, from a careful comparison
with experiments we conclude that at present we cannot safely reject the 
model on the grounds of published CMB anisotropy data. 
If bias is close to one on large scales, galaxy correlation data rules
out the models. New, very stringent constraints come from peculiar
velocities.

Investigating the large-$N$ limit, we argue that our main conclusions
apply to all global $O(N)$ models of structure formation.
\end{abstract}
%\pacs{PACS: 98.80-k, 98.80Cq, 05.40+j}

\section{Introduction}
Recently, a lot of effort has gone into the
determination of cosmological parameters from measurements of cosmic
microwave background (CMB) anisotropies, especially in view of the two
planned satellite experiments MAP and PLANCK\cite{PLMAP}. 
However, we believe it is important
to be aware of the heavy modeling which enters these results. In
general,  simple power law initial spectra for scalar
and tensor perturbations and vanishing vector perturbations are  assumed, as
predicted from inflation. To reproduce observational data,
the composition of the dark matter and the cosmological
parameters  as well as the input spectrum and the scalar to tensor
ratio are varied\cite{infla}. 

In this work, we assume that cosmic structure was induced by scaling
seeds. We
follow the philosophy of a general analysis of scaling seed models
motivated in Ref.~\cite{DK}. 

 Seeds are an
inhomogeneously distributed form of matter (like {\em e.g.} topological
defects) which interacts with the cosmic fluid only
gravitationally and which represents always a small fraction of the
total energy of the universe. They induce geometrical
perturbations, but their influence on the evolution of the background
universe can be neglected. Furthermore, in first order perturbation 
theory, seeds evolve according to the unperturbed spacetime geometry. 

Here, we mainly investigate the models of structure formation with global
texture. This models (for $\Om_{matter}=1$) show discrepancies with
the observed intermediate scale CMB anisotropies and with the galaxy
power spectrum on large scales\cite{PST}. Recently it has been argued 
that the addition of a
cosmological constant leads to better agreement with data for the 
cosmic string model of structure formation~\cite{Avel}. We analyze 
this question for the
texture model, by using ab initio simulation of cosmic texture as
described in Ref.~\cite{ZD}. We determine the CMB anisotropies, the
dark matter power spectrum and the bulk velocities for these models.
We also compare our results with the large-$N$
limit of global $O(N)$ models, and we discuss briefly which type of parameter
changes in the $2$-point functions of the seeds may lead to better 
agreement with data.

We find that the absence of significant acoustic peaks in the CMB
anisotropy spectrum is a robust result for global texture as well as
for the large-$N$ limit for all choices of cosmological parameters
investigated. Furthermore, the dark matter power spectrum on large
scales, $\la\gsim 20h^{-1}\Mpc$, is substantially lower than the
measured galaxy power spectrum.

However, comparing our CMB anisotropy spectra with present data,
we cannot safely reject the model.
On large angular scales, the CMB spectrum is in quite good
agreement with the COBE data set,  while on smaller scales 
we find a significant disagreement only with the Saskatoon
experiment. Furthermore, for non-satellite experiments foreground contamination
remains a serious problem due to the  limited sky and frequency coverage. 

The dark matter power spectra are clearly too low on large scales, but in
view of the unresolved biasing problem, we feel reluctant to rule out
the models on these grounds. A much clearer rejection may come from the
bulk velocity on large scales. Our prediction is by a factor 3 to 5 
lower than the POTENT result on large scales.

Since global texture and the large-$N$ limit lead to very similar
results, we conclude that all global $O(N)$ models of
structure formation for the cosmogonies investigated in this work
are ruled out if the bulk velocity on scales of $50h^{-1}$Mpc is
around 300km/s or if the CMB primordial anisotropies power spectrum
really shows a structure of peaks on sub-degree angular scales.

The formalism used for our calculations  is not presented here but can
be found in\cite{lam}. There, we also explain in detail
the  eigenvector expansion which allows to calculate the CMB
anisotropies and matter power spectra in models with seeds from the
two point functions of the seeds alone. 
 Section~2 is devoted
to a brief description of the numerical simulations. In Section~3 we 
analyze our results and in Section~4 we draw some conclusions. 
\vspace{0.1cm}\\
{\bf Notation:}\hspace{3mm}
We always work in a spatially flat Friedmann universe. The metric is 
given by 
\[ds^2=a(t)^2(dt^2-\de_{ij}dx^idx^j)~,\]
 where $t$ denotes conformal time.

Greek indices denote spacetime coordinates $(0$ to $3)$ whereas Latin
ones run from 1 to 3. Three dimensional vectors are denoted by bold
face characters.

\section{The numerical simulations}

Like in~\cite{ZD}, we consider a spontaneously broken $N$-component scalar
field with O($N$) symmetry. We use the  $\si$-model
approximation, {\em i.e.}, the equation of motion
\be
\Box\b-\l(\b\cd\Box\b\r)\b=0, \label{sigma}
\ee
where $\b$ is the rescaled field $\b=\phi/\eta$.

We do not solve the equation of motion directly, but use a discretized
version of the action~\cite{PST2}:
\be
S=\int d^4\!x \,a^2(t) \l[\frac{1}{2} \dd_\mu \b \cd \dd^\mu \b
	+\frac{\la}{2} \l(\b^2-1\r)\r]\quad,
\ee
where $\la$ is a Lagrange multiplier which fixes the field to the
vacuum manifold (this corresponds to an infinite Higgs
mass). Tests have shown that this formalism agrees
well with the complementary approach of using the equation of
motion of a scalar field with Mexican hat potential and setting
the inverse mass of the particle to the smallest scale that can
be resolved in the simulation (typically of the order of $10^{-35}$ GeV),
but tends to give better energy momentum conservation.

As we cannot trace the field evolution from the unbroken phase
through the phase transition due to the limited dynamical range,
we choose initially a random field at a comoving time $t=2\De x$.
Different grid points are uncorrelated at all earlier
times~\cite{aberna}.

\begin{figure}[ht]
\centerline{\epsfig{figure=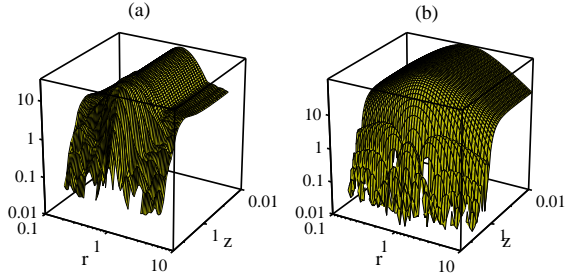,width=7.5cm}}
\caption{\label{fig1}The two point correlation function 
 $C_{11}(z,r) = 
 k^4\sqrt{tt'}\langle\Phi_s(\bk,t)\Phi_s^*(\bk,t')\rangle$ is
 shown. Panel (a) represents the result from numerical simulations of
 the texture model; panel (b) shows the large-$N$ limit.  For
 fixed $r$ the correlator is constant for $z<1$ and then decays. Note
 also the symmetry under $r\ra 1/r$.}
\end{figure}

\begin{figure}[ht]
\centerline{\psfig{figure=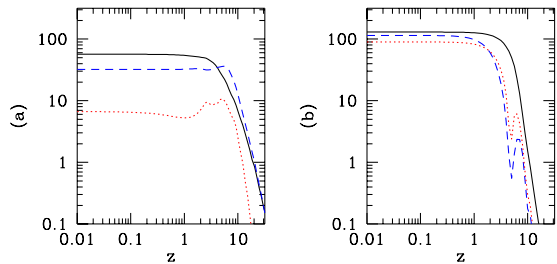,width=7.5cm}}
\caption{\label{fig2}
 The correlators  $C_{ij}(z,1)$ are
 shown. The solid, dashed and dotted lines represent $C_{22}~,~C_{11}$
 and $|C_{12}|$ respectively. Panel (a) is obtained from numerical
 simulations of the texture model and panel (b) shows the large-$N$
 limit. A striking difference is that the large-$N$ value for
 $|C_{12}|$ is relatively well approximated by the perfectly coherent
 result $\sqrt{|C_{11}C_{22}|}$ while the texture curve for $|C_{12}|$
 lies nearly a factor 10 lower.}
\end{figure}

\begin{figure}[ht]
\centerline{\psfig{figure=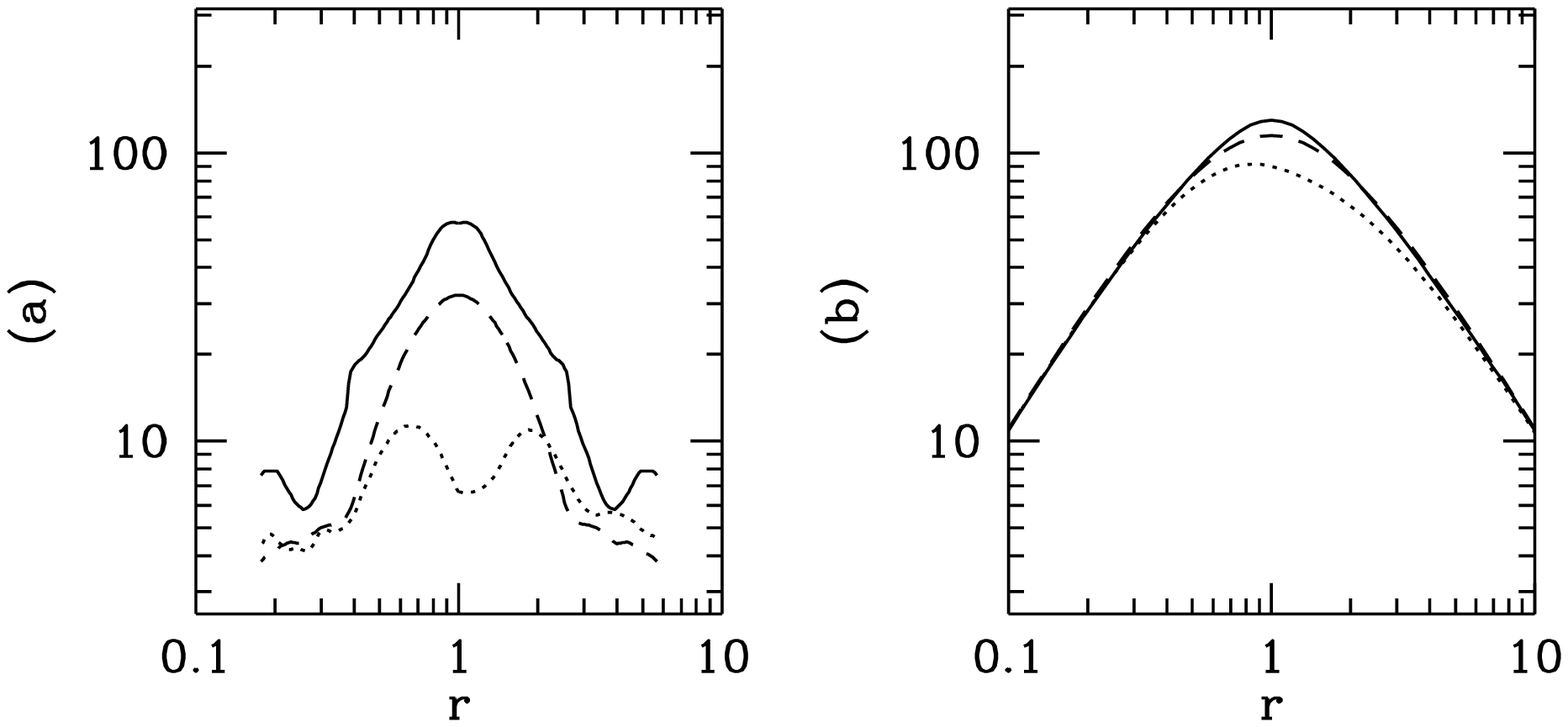,width=7.5cm}}
\caption{\label{fig3}
The correlators  $C_{ij}(0,r)$ are shown in the same line styles as in
Fig.~\ref{fig2}, but for $z=0$ as function of $r=t'/t$. 
The stronger decoherence of the texture model is even
more evident here.}
\end{figure}

To determine the CMB anisotropies and dark matter power spectra, we need
to calculate the unequal time correlators (UTC) of the energy momentum
 tensor or the gravitational field induced by the seeds. 
In Figs~\ref{fig1},~\ref{fig2},~\ref{fig3}, we present as an example
one of the scalar UTC's and the scalar time correlators, 
$r=t'/t=1$, as well as the values
at $z=(k^2tt')=0$ as a function of $r$. Since there are two scalar
degrees of freedom, the two Bardeen potentials $\Phi$ and $\Psi$ there
are three correlators:
\bea
C_{11}(z,r) &=&  k^4\sqrt{tt'}\langle\Phi(k,t)\Phi^*(k,t')\rangle \\
C_{22}(z,r) &=&  k^4\sqrt{tt'}\langle\Psi(k,t)\Psi^*(k,t')\rangle \\
C_{11}(z,r) &=&  k^4\sqrt{tt'}\langle\Phi(k,t)\Psi^*(k,t')\rangle~. 
\eea
% FIGURE 1
% FIGURE 4
% FIGURE 5

The  UTCs are obtained numerically as functions of the 
variables $k$, $t$ and $t_c$ with $t\geq t_c$ and $t_c$ fixed. They 
are then linearly interpolated to the required range. We
construct a hermitian $100\times 100$ matrix in $kt$ and $kt'$, 
with the values of $kt$ chosen on a linear scale to maximize the 
information content, $0\leq kt\leq x_{\mathrm max}$. The choice of 
a linear scale ensures good convergence
of the sum of the eigenvectors after diagonalization 
(see~ \cite{lam}), but still retains
enough data points in the critical region, ${\cal O}(x) =1$,
 where the correlators
start to decay. In practice we choose as the endpoint $x_{\mathrm max}$
of the range sampled by the simulation the value at which the correlator
decays by about two orders of magnitude, typically $x_{\mathrm max}\approx 40$.
The eigenvectors that are fed into a Boltzmann code
are then interpolated using cubic splines with the condition
$v_n(kt)\ra 0$ for $kt \gg x_{\mathrm max}$.

\begin{figure}
\centerline{\psfig{figure=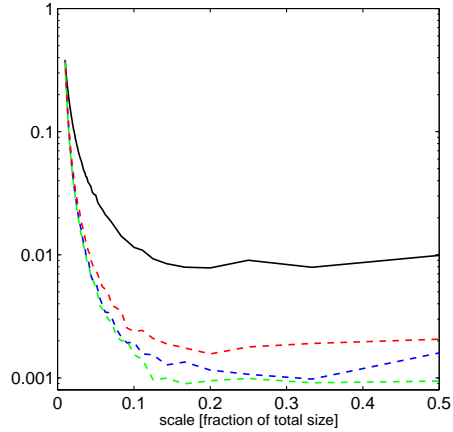,width=6.5cm}}
\caption{\label{fig21}
Energy momentum conservation of our numerical simulations
is shown. The lines represent the sum of the terms which has to vanish
if energy (solid) respectively momentum (dashed) is conserved,
divided by the sum of the absolute value of these terms. The abscissa
indicates the wavelength of the perturbation as fraction of the size
of the entire grid. }
\end{figure}

We use several methods to test the accuracy of the simulation: 
energy momentum conservation of the defects code is found to be
better than 10\% on all scales larger than about 4 grid units, as is 
seen in Fig.~\ref{fig21}. A comparison with the exact
spherically symmetric solution in non-expanding space \cite{PST2}
shows very good agreement. 
% FIGURE 17

\begin{figure}
\centerline{\psfig{figure=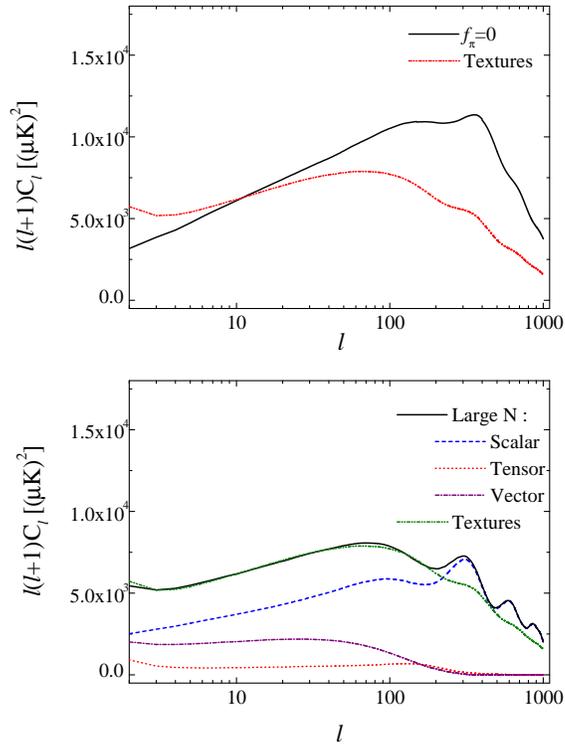,width=7.5cm}}
\caption{\label{fig19}
Top panel: the ${\it f_{\pi}}=0$ model.
Bottom panel: The $C_\ell$ power spectrum is shown for the large-$N$ limit
(bold line) and for the texture model. The main difference is
clearly that the large-$N$ curve shows some acoustic oscillations
which are nearly entirely washed out in the texture case.}
\end{figure}

The resulting CMB spectrum on Sachs Wolfe scales is  consistent with 
the line of sight integration of Ref.~\cite{ZD}.
Furthermore, the overall shape and amplitude of the unequal time correlators
are quite similar to those found in the analytic large-$N$ approximation
\cite{TS,KD,DK}  (see Figs.~1,2 and 3). The main difference of the 
large-$N$ approximation is
that there the field evolution, Eq.~(\ref{sigma}), is approximated by a
linear equation. The non-linearities in the large-$N$ seeds which are
due solely to the energy momentum tensor being quadratic in the
fields, are much weaker than in the texture model where the field
evolution itself is non-linear. Therefore, decoherence which is a
purely non-linear effect, is expected to be much weaker in the
large-$N$ limit. This is actually the main difference between the two
models as can be seen in Fig.~\ref{fig19}.
% FIGURE 18

\section{ Results and comparison with data}

\subsection{CMB anisotropies}

\begin{figure}[htb]
\centerline{\psfig{figure=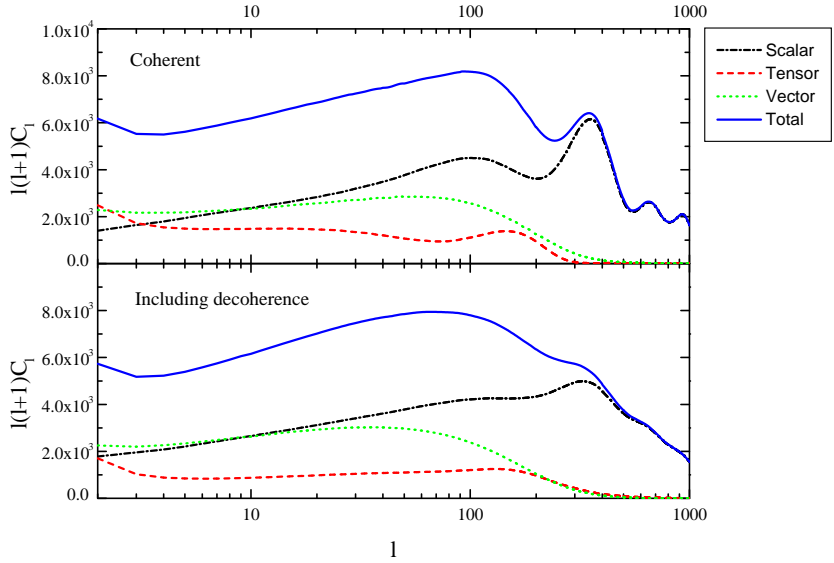,width=7.5cm,angle=-90}}
\caption{\label{fig17}
The $C_\ell$ power spectrum for the texture scenario is shown
in the perfectly coherent approximation (top panel) and in the full
eigenfunction expansion. Even in the coherent approximation, the
acoustic peaks are not higher than the Sachs Wolfe
plateau. Decoherence (see~\protect\cite{lam}) just washes out the 
structure but does not significantly damp the peaks.}
\end{figure}

The $C_\ell$'s for the 'standard' global texture model are shown in 
Fig.~\ref{fig17} (bottom panel), where a comparison of 
the full result with the
totally coherent approximation (see~\cite{lam}) is presented.
% FIGURE 14

% FIGURE 19
\begin{figure}
\centerline{\psfig{figure=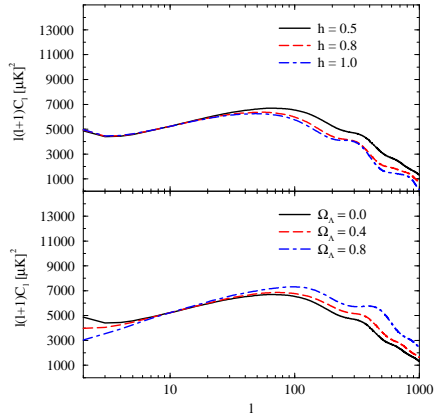,width=7cm}}
\caption{\label{fig23}
The $C_\ell$ power spectrum is shown for different values of
cosmological parameters. In the top panel we choose $\Om_\La=0$,
$\Om_{CDM}=0.95$, $\Om_b=0.05$ and vary $h$. In the bottom panel we
fix $h=0.5$, $\Om_b=0.05$ and vary  $\Om_\La$. We only consider
spatially flat universes, $\Om_0=1$.}
\end{figure}

\begin{table}[ht]
\begin{tabular}{||c|c|c|c||}
\hspace*{1.2cm}$\Om_\La$\hspace{1.2cm} & \hspace{1.2cm}$h$ \hspace{1.2cm}&
$\ep$ & \hspace{1.2cm}$\si_8$\hspace*{1.2cm}\\
\hline
 0.0 & 0.5 &\hspace{0.4cm} $(1.66 \pm 0.17) 10^{-5}$ \hspace{0.4cm}& 0.24 \\
 0.0 & 0.8 & $(1.67 \pm 0.17) 10^{-5}$ & 0.34 \\
 0.0 & 1.0 & $(1.68 \pm 0.17) 10^{-5}$ & 0.44 \\
 0.4 & 0.5 & $(1.64 \pm 0.16) 10^{-5}$ & 0.22 \\
 0.8 & 0.5 & $(1.59 \pm 0.16) 10^{-5}$ & 0.16  \\
\hline
\end{tabular}
\caption{\label{tabeps}
The value of the normalization constant $\ep$ and the fluctuation
amplitude $\si_8$ are given for
the different models considered. The error in $\ep$ comes
from a best fit normalization to the full CMB data set.
 Cosmological parameters which are not
indicated are identical in all models or given by
$\Om_0=\Om_{cdm}+\Om_\La+\Om_b=1$. We consider only spatially flat
models with $\Om_b=0.05$ and a helium fraction of 23\%. The parameter
choice indicated in the top line is referred to as {\em standard}
texture model in the text.}
\end{table} 

Vector and tensor modes are found to be of the same order as 
the scalar component at COBE-scales.
 For the 'standard' texture model 
we obtain $C_{10}^{(S)} : C_{10}^{(V)} : C_{10}^{(T)} \sim  0.9 : 1.0 : 0.3$,
in good agreement with the predictions of Refs.~\cite{PST,Aetal,ABR} 
and \cite{DK}.
Due to tensor and vector contributions, even assuming perfect coherence 
(see Fig.~\ref{fig17}, top panel), the total power spectrum does not  
increase from large to small scales.
Decoherence leads to smoothing of  oscillations in the power
spectrum at small scales and the final power spectrum has a smooth
shape with a broad, low isocurvature 'hump' at $\ell\sim 100$ and a
small residual of the first acoustic peak at $\ell \sim 350$.
There is no structure of peaks at small scales. The power
spectrum is well fitted by the following fourth-order 
polynomial in $x= \log \ell$:

\be {\ell ( \ell +1 ) C_{\ell}\over 110C_{10}} = 1.5 -2.6 x +
3.3 x^2 -1.4 x^3 +0.17x^4 ~.\ee

The effect of decoherence is less important
for the large-$N$ model, where oscillations and peaks are still visible
 (see Fig~\ref{fig19}, bottom panel). This is due to the fact that 
the non-linearity of the large-$N$ limit is  only in the quadratic 
energy momentum tensor. The scalar field evolution is linear in 
this limit\cite{TS}, in contrast to
the $N=4$ texture model. Since decoherence is inherently due to
non-linearities, we expect it to be stronger for lower values of $N$. 
COBE normalization leads to $\epsilon = (0.92 \pm 0.1) 10^{-5}$
 for the large-$N$ limit.

In Fig.~\ref{fig23} we plot the global texture $C_{\ell}$ power spectrum 
for different choices of cosmological parameters. The  variation of 
parameters leads to  similar effects like in the inflationary case,
but with  smaller amplitude.
At small scales ($\ell \ge 200$), the $C_\ell$s decrease with
increasing $H_0$ and they increase when a cosmological
constant $\Omega_{\Lambda} = 1 - \Omega_m$ is introduced.
Nonetheless, the amplitude of the anisotropy power spectrum at
high $\ell$s remains in all cases on the same level like the one at
low  $\ell$s, without showing the  peak found in 
inflationary models.
The absence of acoustic peaks is a stable prediction
of global $O(N)$ models.
 The models are normalized to the full CMB data set,
 which leads to slightly larger values of the normalization 
parameter $\ep=4\pi G\eta^2$ than pure COBE normalization.
In Table~\ref{tabeps} we give the cosmological parameters and the value of
$\ep$ for the models shown in Fig.~\ref{fig23}.

\begin{figure}
\centerline{\psfig{figure=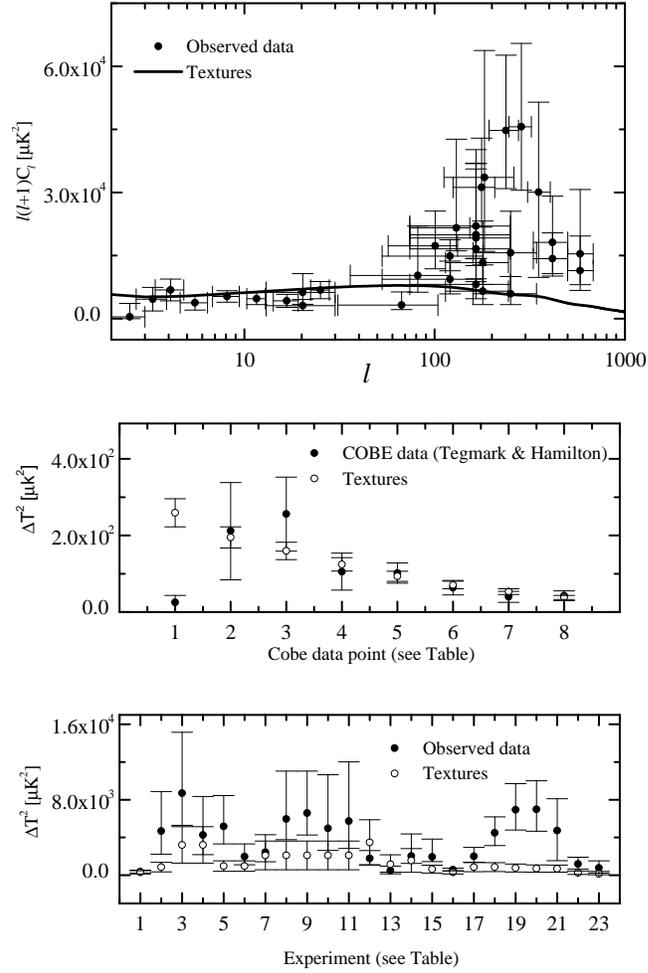,width=9.5cm}}
\caption{\label{fig24}
The $C_\ell$ spectrum obtained in the standard texture model
is compared with data. In the top panel experimental results and the
theoretical curve are shown as functions of $\ell$. In the two lower
panels we indicate the value of each of the 31 experimental data
points with 1-$\si$ error bars  and the  corresponding theoretical value
with its uncertainty. The experiments corresponding to a given number are
given in Table~\ref{tabdata}. In the middle panel the 8 COBE data
points are shown. In the bottom panel other experiments are presented.}
\end{figure}

In order to compare our results with current experimental data, we have
selected a set of $31$ different anisotropy
detections obtained by different experiments, or by the same
experiment with different window functions and/or at different
frequencies. Theoretical predictions and data of CMB anisotropies are 
usually compared by plotting the 
theoretical $C_{\ell}$ curve along with the CMB measurements
converted to band power estimates. We do this in the top panel
of Fig.~\ref{fig24}. The data points  raise
from large to smaller scales, in 
contrast to the theoretical predictions of the model.
This fashion of presenting the data is surely correct, but
lacks informations about the uncertainties in the theoretical model.
Therefore we also compare the detected mean square anisotropy,
 $\Delta^{(Exp)}$ and the experimental 1-$\sigma$ error,
$\Sigma^{(Exp)2}$,  directly 
with the corresponding theoretical mean square anisotropy, given by
\be 
\Delta^{(Th)} = {1 \over {4 \pi}} \sum_{\ell} (2 \ell +1) C_{\ell}
W_{\ell}~,
\ee
where the window function $W_{\ell}$ contains all experimental details
(chop, modulation, beam, etc.) of the experiment.

% FIGURE 20

The theoretical error in principle depends on the statistics of the 
perturbations. If the  distribution is Gaussian, one can associate a 
sample/cosmic variance
\be \Sigma^{(Th)2}={1 \over f}{1\over 8 \pi^2} \sum_{\ell}
(2 \ell +1)W^2_{\ell}C^2_{\ell}~,
  \label{Gauss}\ee
where $f$ represents the fraction of the sky sampled by a given experiment.

 Deviation from Gaussianity leads to an enhancement of this
variance, which can be as large as a factor of $7$ (see \cite{xiao}).
Even if the perturbations are close to Gaussian (which has
been found by simulations on large scales~\cite{ZD,ACSSV}), 
the $C_\ell$'s, which are the squares of Gaussian variables, are 
non-Gaussian. This effect is, however only relevant for relatively low
$\ell$s.
Keeping this caveat in mind, and missing a more precise alternative, we
nevertheless indicate the minimal, Gaussian error calculated
according to~(\ref{Gauss}).
We add a $30 \%$ error from the CMB normalization. The numerical
seeds are assumed to be about 10\% accurate. 

In Table~\ref{tabdata}, the detected mean square anisotropy,
 $\Delta^{(Exp)}$, with the experimental 1-$\sigma$ error 
 are listed for each experiment of our data set. The
 corresponding sky coverage is also indicated.
 In Fig.~\ref{fig24} we plot these data points, together with the 
 theoretical predictions for a texture model with $h=0.5$ and 
 $\Omega_{\Lambda} = 0$.

\begin{table}
\footnotesize
\begin{center}
\begin{tabular}{||c|c|c|c|c|c|c|c||}
\hline\hline
Experiment & Data point   &  $\Delta T^2 (\mu K)^2$ & 
$ + (\mu K)^2$ & 
$ - (\mu K)^2 $ &Sky Coverage 
& Reference& $\chi^2_j$ \\ 
\hline
COBE1 & 1& 25.2 & 183 & 25.2 & 0.65 & \protect\cite{tegma}& 125.29 \\
COBE2 & 2& 212 & 126 & 128 & 0.65 & \protect\cite{tegma}& 0.02 \\
COBE3 & 3& 256 & 96.5& 96.9& 0.65& \protect\cite{tegma}& 0.49 \\ 
COBE4 & 4& 105.5 & 48.3&48.2 & 0.65& \protect\cite{tegma}& 0.74\\
COBE5 & 5& 101.9 & 26.5& 26.4& 0.65& \protect\cite{tegma}& 0.1\\ 
COBE6 & 6& 63.4 & 19.11& 18.9& 0.65& \protect\cite{tegma}& 1.11\\ 
COBE7 & 7& 39.6 & 14.5 & 14.5& 0.65&\protect\cite{tegma} & 2.55\\ 
COBE8 & 8& 42.5 &12.7& 12.8& 0.65&\protect\cite{tegma} & 0.04\\ \hline
ARGO Hercules& 1& 360& 170& 140& 0.0024&\protect\cite{deb}& 0.001\\ \hline
MSAM93 & 2&4680&4200&2450&0.0007& \protect\cite{che}& 0.74\\ \hline
MSAM94 & 3& 4261 & 4091 & 2087  & 0.0007& \protect\cite{che2}&0.51\\
MSAM94 & 4& 1960 & 1352 &  858  & 0.0007& \protect\cite{che2} &0.01\\ \hline
MSAM95 & 5& 8698 & 6457 & 3406  & 0.0007& \protect\cite{che3} &1.47\\
MSAM95 & 6& 5177 & 3264 & 1864  & 0.0007& \protect\cite{che3} &0.30\\ \hline
MAX HR& 7& 2430& 1850&1020&0.0002&\protect\cite{tan}&0.001\\
MAX PH& 8& 5960& 5080&2190&0.0002&\protect\cite{tan}&0.41\\ 
MAX GUM& 9& 6580& 4450&2320&0.0002&\protect\cite{tan}&0.73\\ 
MAX ID& 10& 4960& 5690&2330&0.0002&\protect\cite{tan}&0.17 \\ 
MAX SH& 11& 5740& 6280&2900&0.0002&\protect\cite{tan}& 0.25\\ \hline
Tenerife & 12& 3975 & 2855 & 1807 & 0.0124& \protect\cite{gut} & 0.64\\ \hline
South Pole Q& 13& 480& 470& 160 & 0.005& \protect\cite{gun} &0.52 \\ 
South Pole K& 14& 2040& 2330& 790 & 0.005& \protect\cite{gun} & 0.01\\ \hline
Python& 15& 1940& 189& 490& 0.0006& \protect\cite{dra} & 0.37 \\ \hline
ARGO Aries& 16& 580& 150& 130& 0.0024& \protect\cite{mas}& 0.78\\ \hline
Saskatoon&17&1990&950&630&0.0037&\protect\cite{net}&0.79\\ 
Saskatoon&18&4490&1690 & 1360&0.0037&\protect\cite{net}&3.83\\
Saskatoon&19&6930&2770&2140&0.0037&\protect\cite{net}&4.60\\ 
Saskatoon&20&6980&3030&2310&0.0037&\protect\cite{net}&4.01\\ 
Saskatoon&21&4730&3380&3190&0.0037&\protect\cite{net}&1.32\\ \hline
CAT1 & 22& 934 & 403 & 232  & 0.0001&\protect\cite{bak} &1.36\\
CAT2 & 23& 577 & 416 & 238  & 0.0001&\protect\cite{bak} &0.62\\ \hline
\hline
\end{tabular}
\end{center}
\vspace{0.3cm}
\caption{The CMB anisotropy detections used in our analysis. The 3.,
4. and 5. column denote the value of the anisotropy and the upper and
lower 1-$\si$ errors respectively.
The references are: 
Tegmark and Hamilton 1997 \protect\cite{tegma}; de Bernardis {\it et
al.} 1994 \protect\cite{deb}; Cheng {\it et al.} 1994 \protect\cite{che}; 
Cheng {\it et al.} 1996
 \protect\cite{che2};  Cheng {\it et al.} 1997 \protect\cite{che3}; 
 Tanaka {\it et al.} 1996 \protect\cite{tan}; Gutierrez {\it et al.} 1997 
\protect\cite{gut}; Gundersen {\it et al.} 1993 \protect\cite{gun}; 
 Dragovan {\it et al.} 1993 \protect\cite{dra}; Masi et al 1996
\protect\cite{mas}; Netterfield {\it et al.} 1996 \protect\cite{net};
 Scott {\it et al.} 1997 \protect\cite{bak}.
 \label{tabdata}}
\end{table}
 
We find that, apart from the
COBE quadrupole, only the Saskatoon experiment disagrees
significantly, more than $1\si$, with our model. But also this
disagreement is below $3\si$ and thus not sufficient to rule
out the model. 
In the last column of Table~\ref{tabdata} we indicate
\[ \chi_j^2= (\De_j^{(Th)}-\De_j^{(Exp)})^2/(\Si_j^{(Th)2}+\Si_j^{(Exp)2}) \]
for the $j$-th experiment, where the theoretical model is the standard
texture model with
$\Omega_{\Lambda}=0$ and $h=0.5$. The major discrepancy
between data and theory comes from the COBE quadrupole.
Leaving away the quadrupole, which can be contaminated 
and leads to a similar $\chi^2$ also for  inflationary models, 
the data agrees quite well with the model, with the exception of 
three Saskatoon  data points.
Making a rough chi-square analysis, we obtain
(excluding the quadrupole) a value $\chi^2=\sum_j\chi_j^2 \sim 30$ for
a total of 30 data points and one constraint. An absolutely reasonable
value, but one should take into account that
the experimental data points which we are considering
are not fully independent. The regions of sky sampled by the
Saskatoon and MSAM  or COBE and Tenerife,
 for instance, overlap.
Nonetheless, even reducing the degrees of freedom of our analysis
to $N = 25$, our $\chi ^2$ is still in the range $(N-1) \pm {
\sqrt 2 (N-1)} \sim 24 \pm 7$ and hence still compatible with 
the data.

This shows that even assuming Gaussian statistics,
 the models are not convincingly ruled out from present CMB data. 
There is however one caveat in this analysis: A chi-square test is not
sensitive to the sign of the discrepancy between theory and
experiment. For our models the theoretical curve is systematically 
lower than the experiments. For example, whenever the discrepancy between
theory and data is larger than $0.5\si$, which happens with nearly half
of the data points (13), in all cases except for the COBE quadrupole, 
the theoretical value is smaller than the data. If
smaller and larger are equally likely, the probability to have $12$ or
 more equal signs  is $2(13+1)/2^{13}\simeq 3.4\times 10^{-3}$. This  
indicates that either the model
is too  low or that the data points are systematically too high. The
number  $0.003$ can however not be taken seriously, because we can
easily change it by  increasing our normalization on a moderate cost of
$\chi^2$.

\subsection{Matter distribution}

In Table~\ref{tabeps} we show the expected variance 
of the total mass 
fluctuation $\sigma_R$ in a ball of radius $R=8h^{-1}$Mpc,
for different choices of cosmological parameters.
We find $\sigma_8 = (0.44 \pm 0.07)h$
(the error coming from the CMB normalization) for a flat
model without cosmological constant, in agreement
with the results of Ref.~\cite{PST}.
From the observed  cluster abundance, one infers 
$\sigma_8 = (0.50 \pm 0.04) \Omega^{-0.5}$ \cite{eke} and
$\sigma_8 = 0.59^{+0.21}_{-0.16}$ \cite{lid}.  These results,
 which are obtained with the Press-Schechter formula, assume Gaussian 
 statistics. We thus have to take them with a
grain of salt, since we do not know how non-Gaussian fluctuations on
cluster scales are in the texture model. According to Ref.~\cite{free}, the 
Hubble constant lies in the interval
$h\simeq 0.73 \pm 0.06 \pm 0.08$. Hence,
in a flat CDM cosmology, taking into account the  uncertainty of
the Hubble constant, the texture scenario predicts a reasonably consistent 
value of $\sigma_8$.

As already noticed in Refs.~\cite{ABR} and \cite{PST},
unbiased global texture models are unable to reproduce
the power of galaxy clustering at very large scales,
 $\gsim 20 h^{-1}$ Mpc.
In order to quantify this discrepancy we compare our prediction of the linear
matter power spectrum with the results from a number of infrared 
(\cite{fish},\cite{tadr}) and optically-selected (\cite{daco}, \cite{lin})
galaxy redshift surveys, and with the real-space power
spectrum inferred from the APM photometric sample (\cite{baugh}) (see 
Fig.~\ref{fig26}). Here, cosmological parameters have important
effects on the shape and amplitude of the matter power spectrum.
Increasing the Hubble constant shifts the peak of the power spectrum 
to smaller scales (in units of $h/$Mpc), while the inclusion of a cosmological
constant enhances large scale power. 

 We consider a set of models in 
 $\Omega_{\Lambda}$ -- $h$ space, with linear bias \cite{kais} as
additional parameter. In  Table~\ref{tabpow} we report
  for each survey and for each model the best value of the bias
parameter obtained by $\chi^2$-minimization. We also indicate the
value of $\chi^2$ (not divided by the number of data points).
The data points and the theoretical predictions are plotted in 
Fig.~21.
Our bias parameter  
strongly depends on the data considered. This is not surprising, since
also  the catalogs are biased relative to each other.

Models without cosmological constant and with $h \sim 0.8$ only require
a relatively modest bias $b \sim 1.3 -3$. But for these models the
shape of the power spectrum is wrong as can be seen from the
value of  $\chi^2$ which is much too large.
The bias factor is in agreement with our prediction for $\sigma_8$.
 For example, our best fit for the IRAS data, for 
 $h \sim 0.8$ is $b \sim 1.3$. With $\sigma_8^{IRAS} = 
(0.69 \pm 0.05)$, this gives  $\sigma_8 \sim 0.48 \pm 0.04$,
compatible with the direct computation

Whether IRAS galaxies are biased  is still under debate. Published values for
the $\beta$ parameter, defined as $\beta= \Omega^{0.6}/b$,
for IRAS galaxies, range between
$\beta_I = 0.9^{+0.2}_{-0.15}$ \cite{kais2} and $\beta_I = 0.5 \pm
0.1$ \cite{will}.  Biasing of IRAS galaxies is
also suggested by measurements of bias in the
optical band. For example, Ref.~\cite{pea97} finds
 $\beta_o =0.40 \pm 0.12$, in marginal agreement 
 with \cite{shai}, which obtains $\beta_o = 0.35 \pm 0.1$.
A bias for IRAS galaxies is not only possible but
even preferred in {\em flat} global texture models. 

But also with  bias, our models are in significant contradiction
with the shape of the power spectrum at large scales.
As the values of $\chi^2$ in Table~\ref{tabpow} and Fig.~\ref{fig26}
 clearly indicate, the models are inconsistent with the shape of the IRAS
power spectrum, and they can be rejected with a high confidence
 level. The APM data which  has the smallest error bars is the most
 stringent evidence against texture models. Nonetheless, these data points
 are not measured in redshift space but they 
 come from a de-projection of a $2-D$ catalog into $3-D$ space.
 This might introduce systematic errors and thus the errors of APM may
be underestimated.

\begin{table}
\begin{tabular}{||c|c|c|c|c|c||}
\hline 
Catalog & \hspace{0.3cm}$h$ \hspace{0.3cm} & \hspace{0.3cm}
  $\Omega_{\Lambda}$\hspace{0.3cm} & Best fit bias $b$& $\chi^2$&Data points\\
\hline
 CfA2-SSRS2 101 Mpc& 0.5 & 0.0 & 3.4 & 29 & 24\\
 CfA2-SSRS2 101 Mpc& 0.8 & 0.0 & 2.0 & 40 & 24\\
 CfA2-SSRS2 101 Mpc& 1.0 & 0.0 & 1.9 & 44 & 24\\
 CfA2-SSRS2 101 Mpc& 0.5 & 0.4 & 3.9 & 17 & 24\\
 CfA2-SSRS2 101 Mpc& 0.5 & 0.8 & 9.5 &  4 & 24\\
\hline
 CfA2-SSRS2 130 Mpc& 0.5 & 0.0 & 5.3 &  8 & 19\\
 CfA2-SSRS2 130 Mpc& 0.8 & 0.0 & 3.4 & 15 & 19\\
 CfA2-SSRS2 130 Mpc& 1.0 & 0.0 & 3.4 & 16 & 19\\
 CfA2-SSRS2 130 Mpc& 0.5 & 0.4 & 5.6 &  5 & 19\\
 CfA2-SSRS2 130 Mpc& 0.5 & 0.8 & 11.1 &  4 & 19\\
\hline
 LCRS  &0.5 & 0.0 & 3.0 &  71 & 19\\
 LCRS  & 0.8 & 0.0 & 1.8 & 96 & 19\\
 LCRS  & 1.0 & 0.0 & 1.6 & 108 & 19\\
 LCRS  & 0.5 & 0.4 & 3.7 &  33 & 19\\
 LCRS  & 0.5 & 0.8 & 8.7 &  40 & 19\\
\hline
 IRAS  &0.5 & 0.0 & 2.3 &  102 & 11\\
 IRAS  & 0.8 & 0.0 & 1.3 & 131 & 11\\
 IRAS  & 1.0 & 0.0 & 1.3 & 140 & 11\\
 IRAS  & 0.5 & 0.4 & 2.8 &  70 & 11\\
 IRAS  & 0.5 & 0.8 & 6.3 &   9 & 11\\
\hline
 IRAS 1.2 Jy & 0.5 & 0.0 & 4.2 &  56 & 29\\
 IRAS 1.2 Jy & 0.8 & 0.0 & 2.9 & 92 & 29\\
 IRAS 1.2 Jy & 1.0 & 0.0 & 2.9 & 99 & 29\\
 IRAS 1.2 Jy & 0.5 & 0.4 & 4.3 &  39 & 29\\
 IRAS 1.2 Jy & 0.5 & 0.8 & 6.7 &   28 & 29\\ 
\hline
 APM & 0.5 & 0.0 & 3.3 & 1350 & 29\\
 APM & 0.8 & 0.0 & 1.8 & 1500 & 29\\
 APM & 1.0 & 0.0 & 1.7 & 1466 & 29\\
 APM & 0.5 & 0.4 & 3.5 & 1461 & 29\\
 APM & 0.5 & 0.8 & 6.2 & 1500 & 29\\
\hline
 QDOT & 0.5 & 0.0 & 4.3 & 32 & 19\\
 QDOT & 0.8 & 0.0 & 2.9 & 44 & 19\\
 QDOT & 1.0 & 0.0 & 2.9 & 46 & 19\\
 QDOT & 0.5 & 0.4 & 4.3 & 25 & 19\\
 QDOT & 0.5 & 0.8 & 7.3 & 14 & 19\\
\hline
\end{tabular}
\vspace{0.3cm}
\caption{\label{tabpow}
Analysis of the matter power spectrum. In the first column the
catalog is indicated. Cols.~2 and 3 specify the model parameters. In
cols.~4 and 5 we give the bias parameter inferred by $\chi^2$
minimization as well as the value of $\chi^2$. Col.~6 shows the number
of 'independent' data points assumed in the analysis.}
\end{table}
  
Models with a cosmological constant agree much better with the shape
 of the observed power spectra, the value of  $\chi^2$ being low for
 all except the APM data.
But the  values of the bias factors are extremely high for these
 models.  For example, IRAS galaxies should have a bias $b \sim 3 -
 6$,   resulting in $\sigma_8 \le 0.25$, and in a $\beta_{I} \le 0.2$
 which is  too small, even allowing for big variances due to
 non-Gaussian statistics.

The power spectra for the large-$N$ limit and for the coherent
approximation are typically a factor 2 to 3 higher (see Fig.~10), and
the biasing problem is  alleviated for these cases. For
$\Om_\La=0$ we find $\si_8=0.57h $ for the
large-$N$ limit and  $\si_8=0.94h$ for the coherent approximation.
This is no surprise since only one source function, $\Psi_s$, the
analog of the Newtonian potential, seeds dark matter fluctuations and
thus the coherence always enhances the unequal time correlator. The
dark matter Greens function is not oscillating, so this enhancement 
translates directly into the power spectrum. 

\begin{figure}
\centerline{\psfig{figure=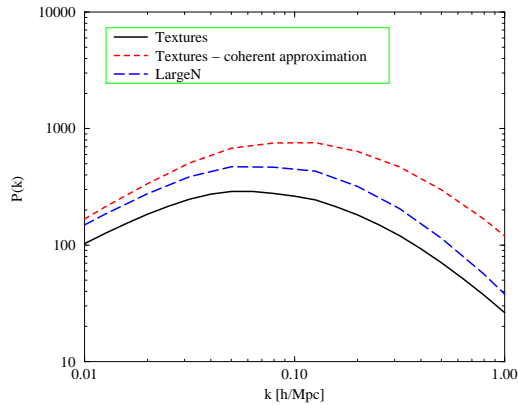,width=7.5cm}}
\caption{\label{fig22}
The dark matter power spectrum for the texture model (solid line) is
compared with the coherent approximation (short dashed) and the
large-$N$ limit (long dashed). The spectra are COBE normalized and the
cosmological parameters are $\Om_\La=0~,~h=0.5$.}
\end{figure}

Models which are anti-coherent in the sense defined in Section~IID
reduce power on Sachs-Wolfe scales and enhance the power in the dark
matter. Anti-coherent scaling seeds are thus the most promising
candidates which may cure some of the problems of global $O(N)$ models.

The simple analysis carried out here does not take into account
the effects of non-linearities and redshift distortions.
Redshift distortions in the texture case should be less important
than in the inflationary case since the peculiar velocities are rather
low (see next paragraph).
 Non-linearities typically set in at $k \ge 0.5h$Mpc$^{-1}$ and should not
have a big effect on our main conclusions which come from much larger
scales.  Inclusion of these corrections will
result in more small-scale power and in a broadening of the spectra,
which even enhances the conflict between models and data.
Furthermore, variations of other cosmological parameters, like
the addition of massive neutrinos, hot dark matter, which is not 
considered here,
will result in a change of the spectrum on small scales but will not
resolve the  discrepancy at large scales.

Nonetheless, scale dependent biasing may exist and 
lead to a non-trivial relation between the calculated dark matter 
power spectrum and the observed galaxy power
spectrum. We are thus very reluctant to rule out the model
by comparing two in principle different things, the relation of which
is far from understood. Therefore we would prefer to reject the models
on the basis of peculiar velocity data, which is more difficult to
measure but most certainly not biased. 

% FIGURE 21
\begin{figure}
\centerline{\psfig{figure=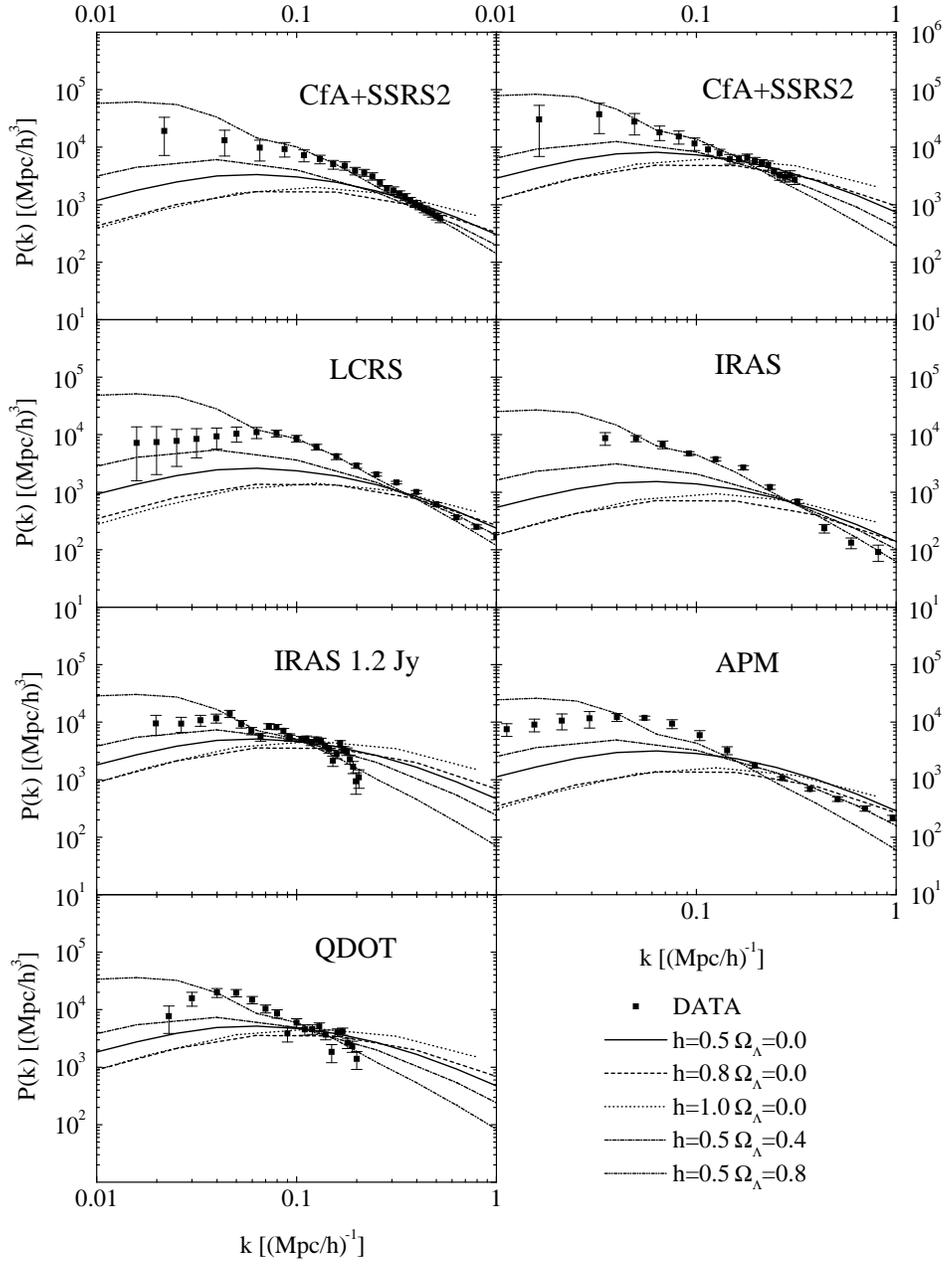,width=12.5cm}}
\caption{\label{fig26}
Matter Power spectrum: comparison between data and theory.
References are in the text. Data set courtesy of M. S. Vogeley 
\protect\cite{voegel}.}
\end{figure}

\subsection{Bulk velocities}

To get a better handle on the missing power on 20 to 100$h^{-1}$Mpc,
 we investigate the
velocity power spectrum which is not plagued by biasing problems. The
assumption that galaxies are fair tracers of the velocity field seems
to us much better justified, than to assume that they are fair tracers
of the mass density. 
We therefore test our models against peculiar velocity data. We
use the data by Ref.~\cite{deke} which gives the bulk flow

\be \si^2_v(R) = {H_0^2\Om_m^{1.2} \over {2 \pi^2}} \int P(k)W(kR)dk ~,\ee

in spheres of radii $R = 10$ to $60 h^{-1}$Mpc. 
These data are derived after reconstructing the $3-$dimensional
velocity field with the POTENT method (see~\cite{deke} and references 
therein).

As we can see from Table~\ref{tabevl}, the COBE normalized texture 
model predicts too low velocities on large scales when compared with
POTENT results.
Recent measurements of the bulk flow lead to somewhat lower estimates
like $\sigma_{v}(R) \sim (230 \pm 90)$ at $R = 60 h^{-1}$Mpc
(\cite{giova}), but still a
discrepancy of about a factor of $2$ in the best case remains.

Including a cosmological constant helps at large scales, but
decreases the velocities on small scales. 

If the observational bulk velocity data is indeed reliable (there
are some doubts about this\cite{Mark}), all global $O(N)$ models are 
ruled out.

\begin{table}
\begin{tabular}{||c|c|c|c|c|c||}
\hline
\hspace{0.7cm}R\hspace{0.7cm} &\hspace{0.7cm} $\sigma_v$ (R)
\hspace{0.7cm}&\hspace{0.7cm} $\Delta_v$\hspace{0.7cm} &\hspace{0.7cm}
 $h=0.5$\hspace{0.7cm}&\hspace{0.7cm} $h=1.0$\hspace{0.7cm}&
 \hspace{0.7cm}$\Omega_{\La}=0.8$\hspace*{0.7cm}\\
\hline
 10 & 494 & 170 & 145 & 205&  86 \\
 20 & 475 & 160 & 100 & 134&  78 \\
 30 & 413 & 150 &  80 &  98&  70 \\
 40 & 369 & 150 &  67 &  78&  65 \\
 50 & 325 & 140 &  57 &  65&  61 \\
 60 & 300 & 140 &  50 &  56&  57 \\
\hline
\end{tabular}
\vspace{0.3cm}
\caption{\label{tabevl}
Bulk velocities: Observational data from \protect\cite{deke} and theoretical
predictions. $\De_v$ estimates the observational uncertainty.
The uncertainties on the theoretical predictions are
around $\sim 30 \%$. The models $\Om_\La=0$ with $h=0.5$ and $h=1$ as well
as $\Om_\La=0.8,~h=0.5$ are investigated.}
\end{table}

\section{Conclusions}
We have determined CMB anisotropies and other power spectra of linear
perturbations in models with global $O(N)$ symmetry which contain global
monopoles and texture. Our main results can be summarized as follows:
\begin{itemize}
\item Global $O(N)$ models predict a flat spectrum
(Harrison-Zeldovich) of CMB anisotropies on large scales which is in
good agreement with the COBE results. Models with vanishing cosmological
constant and a large value of the Hubble parameter give $\si_8\sim
0.4$ to $0.5$ which is reasonable.
\item Independent of cosmological parameters, these models do not
exhibit pronounced acoustic peaks in the CMB power spectrum.
\item The dark matter power spectrum from  global $O(N)$ models with
$\Om_\La=0$ has reasonable amplitude but does not agree in its shape 
with the galaxy power spectrum, especially on very large 
scales $>20h^{-1}$Mpc.
\item Models with considerable cosmological constant agree relatively
well with the shape of the  galaxy power spectrum, but need very high 
 bias $b\sim 4 - 6$ even with respect to IRAS galaxies.
\item  The large scale bulk velocities are by a factor of about 3 to 5 
smaller than the values inferred from \cite{deke}.
\end{itemize}
In view of the still considerable errors in the CMB data (see
Fig.~\ref{fig24}), and the biasing problem for the dark matter power spectrum, 
we consider the last argument as the most convincing one to rule out  
global $O(N)$ models. Even if velocity data is quite uncertain,
observations agree that bulk velocities on the scale of
$50h^{-1}$Mpc are substantially larger than the (50 -- 70)km/s 
obtained in texture models. 

However, all our constraints have been obtained assuming Gaussian
statistics. We know that global defect models are non-Gaussian, but
we have not investigated how severely this influences the above
conclusions. Such a study, which we plan for the future, requires 
detailed maps of fluctuations, the
resolution of which is always limited by computational resources. 
Generically we can just say that non-Gaussianity can only weaken the
above constraints.

Our results naturally lead to the question  whether all scaling seed models
are ruled out by present data.  The main problem of the $O(N)$
model is the missing power at intermediate scales, $\ell\sim 300 -
500$ or $R\sim (20 -100)h^{-1}\Mpc$. We have briefly investigated whether
this problem can be mitigated in a scaling seed model without vector
and  tensor perturbations. In this case, also scalar anisotropic
stresses are reduced by causality requirements (see Ref.~\cite{DK}), and
perturbations on superhorizon scaled are compensated (see~\cite{DS}).
For simplicity, we analyze a model with purely scalar perturbations
and no anisotropic stresses at all, $f_\pi=0$. The seed function
$\Phi_s$ is taken from the texture model (numerical simulations) and
we set $\Psi_s=-\Phi_s$. The resulting CMB anisotropy spectrum is
shown in Fig.~\ref{fig19}, top  panel. A smeared out acoustic peak with
an amplitude of about $2.2$ does indeed appear in this model. This
is mainly due to fluctuations on large scales being smaller, as is
also evident from the higher value of 
$\ep=(2.2\pm 0.2)\times 10^{-5}$. But also here, the dark
matter density fluctuations and bulk velocities are substantially
lower than observed galaxy density fluctuations or the POTENT bulk
flows.

Clearly, this simple example is not sufficient and a more thorough 
analysis of generic scaling seed models is presently
under investigation. So far it is just clear that contributions from
vector and tensor perturbations are severely restricted.
\vspace{0.6cm}\\
{\large\bf Acknowledgment}\\
It is a pleasure to thank Andrea Bernasconi, Paolo de Bernardis,
 Roman Juszkiewicz, Mairi Sakellariadou, Paul Shellard, Andy Yates 
and Marc Davis for stimulating discussions. Our
Boltzmann code is a modification of a code worked out by the group
headed by Nicola Vittorio. We also thank Michael Vogeley who kindly 
provided us the galaxy power spectra shown in our figures.
The numerical simulations have been performed at the Swiss
super computing center CSCS. This work is partially supported by the
Swiss National Science Foundation.

\end{document}